# Making Space for Stories: Ambiguity in the Design of Personal Communication Systems

Paul M. Aoki
Palo Alto Research Center
3333 Coyote Hill Road
Palo Alto, CA 94304-1314 USA

Allison Woodruff
Intel Research Berkeley
2150 Shattuck Avenue, Suite 1300
Berkeley, CA 94704-1347 USA

**ABSTRACT**
Pervasive personal communication technologies offer the potential for important social benefits for individual users, but also the potential for significant social difficulties and costs. In research on face-to-face social interaction, ambiguity is often identified as an important resource for resolving social difficulties. In this paper, we discuss two design cases of personal communication systems, one based on fieldwork of a commercial system and another based on an unrealized design concept. The cases illustrate how user behavior concerning a particular social difficulty, unexplained unresponsiveness, can be influenced by technological issues that result in interactional ambiguity. The cases also highlight the need to balance the utility of ambiguity against the utility of usability and communicative clarity.

**ACM Classification Keywords**
H.4.3 [**Information Systems Applications**]: Communications Applications.

**INTRODUCTION**
Social interaction often requires *face-work* – the measures people take to preserve face for themselves and for others when problematic events occur during interactions [16]. Face-work is used to accomplish goals such as avoiding embarrassment and maintaining harmony in relationships. Face-work involves managing the impressions that other people have of your behavior. In some cases, this simply means ensuring that people have a correct understanding of your actions, while in other cases it may involve giving a misleading impression by telling a "white lie."

For example, imagine that you called your friend two weeks ago and left a message asking them to call you back, but they did not. Based on previous events in your relationship, you account for this failure as likely being due to a busy schedule as opposed to a personal issue. You later run into your friend at a party. Your friend, apparently concerned that you will feel you have been rebuffed, says, "I'm sorry I didn't call you back. I've been out of town." Now, you may or may not believe your friend is telling the truth. However, in either case you are likely to give the impression that you accept your friend's story and move on, saying something like, "That's all right. I just wanted to ask you for Mary's phone number." By taking this action, you are helping to maintain harmony, just as your friend was helping to maintain harmony by offering an explanation. You and your friend have now reached a "working consensus" [15] on a story that allows you to proceed without direct confrontation – your friend is not directly confronted with the suggestion that they were negligent or untruthful, and you are not directly confronted with the suggestion that you have been rebuffed.

Wireless personal communication systems can easily multiply the complexity of such situations, increasing the daily burden of face-work. Because the devices are both portable and personal, people who initiate contact can gain increased insight into one's activities, often achieving access to domains (home, work, friends,…) that one might otherwise try to keep separate [28]. More importantly, the devices support a growing number of lightweight communication services – text messaging, mobile instant messaging, push-to-talk audio messaging, etc. – and thereby enable remote interactions to occur on an increasingly spontaneous basis (see, e.g., [25]). Many communicative acts have implications for the status of a relationship (providing contact information [25], accepting a call [24], replying promptly [22,23], using appropriate language [33]), making each lightweight interaction an opportunity for a problematic event.

Recognizing this, we suggest that the design of personal communication systems can be reconceptualized in a way that helps designers in addressing these issues. Rather than simply looking at novel personal communication systems in terms of their effectiveness in transmitting information, we also need to consider their designs in terms of how they address users' overall social needs. A key observation is that more interpersonal knowledge often makes relationship management harder – sociologists have long argued that "[r]elationships…presuppose a certain ignorance and a measure of mutual concealment" [32] to function smoothly. (In the scenario above, suppose you had perfect information about your friend's movements. If your friend were aware of this, they would be restricted to stories that were consistent with that information; if not, and they told a "white lie," they would be caught.) We might, then, think of evaluating designs in terms of their support for creating personal space through *ambiguity*, a goal that may have to be traded off against the goal of clarity that we usually associate with communication systems.

A central theme of this paper is that, while participants in interaction always observe and account for the actions of others, it is not always desirable for all parties to be able to account accurately and precisely. The communication system, through its design, can provide resources for participants to create ambiguity in their interactions.[1] (Here, and throughout this paper, we use ambiguity in its sense of "admitting of multiple interpretations.") Put another way, rather than trying to provide perfect information to eliminate ambiguity, the designer can try to preserve enough space for the stories that make up the working consensus of successful face-work.

In this paper, we illustrate some ways in which users exploit properties of personal communication technologies – not necessarily consciously – to create ambiguity. In particular, we focus on how people handle two variations of unresponsiveness in interaction: when people fail to respond to others' conversation, or when people fail to reciprocate others' attempts to establish mutual access in a communication medium. In part, we draw upon examples from research literature in mediated communication technologies. However, we mainly discuss implications of our own fieldwork and design experience with a mobile audio communication system (described in part in [2,37]).

**AMBIGUITY AND MEDIATION**
Participants in social interaction attempt to account for others' behavior and to make their own behavior accountable to others; participants continually form meaning from what others say and do, combining this with context such as recent utterances and occurrences, personal knowledge, cultural norms, etc. [12]. In this section, we discuss two main ways in which mediation can affect how people account for each other's behavior. First, we discuss "passive effects" – effects that are natural results of the fact that mediated communication passes through a channel and is therefore less direct than face-to-face communication. Second, we discuss "active effects" – effects that result from explicit efforts to manage the perception of participants. Because much previous literature has discussed the attempts of *humans* to influence how other humans account for their actions, in this discussion of active effects, we focus on *features in the communication system* that are intentionally designed to influence how one participant accounts for actions of another.

Before launching into the discussion of technological aspects, we need to expand upon the relationship between ambiguity and accounting. For a variety of reasons, participants in social interaction cope with substantial ambiguity even in the most ordinary face-to-face encounters. Sometimes, participants deliberately make their actions ambiguous. They may also increase the likelihood of misunderstanding by inadvertently failing to provide enough information about some specific subject. However, the reason why ambiguity is pervasive in interaction is that interaction is full of unstated "common-sense" assumptions that are based on context: participants "just know" how they should behave in public and at home, what kinds of behavior are polite and rude, and so on, and expect others to have analogous knowledge of how to act [12]. Therefore, each participant accounts for others' actions on a continual basis as the interaction unfolds. In short, any discussion of ambiguity in interaction must consider how people account.[2]

**Passive Effects of Mediation**
In this subsection, we describe a framework for thinking about how passive aspects of personal communication technology affect the way in which people account. This framework is intended to be illustrative rather than original, comprehensive or predictive; although it approximates the common assumptions of a variety of writings in social psychology and sociology, it is not derived from or justified by a particular theory.

Consider two people, A and B. In face-to-face interaction (Figure 1, top), person A's *actions* and their subsequent *results*, if any, are directly observable by person B. While person A's *intent* remains ambiguous and must be inferred (since intent cannot be observed), person B has relatively little doubt as to what was done. By contrast, in interaction through a mediated communication system (Figure 1, center and bottom), person A's actions are generally not directly observable, a fact represented here by the metaphorical barrier between A and B. Therefore, in the mediated case person B must make inferences about person A's actions as well as inferences about person A's intent.

If the mediating technology is very simple (or very well-understood by person B), there may only be a trivially small number of plausible actions that explain what person B is observing (Figure 1, center). In this case, B's observation-action inference is easy. For example, say A and B are communicating by telephone (using an ideal service with reliable connectivity). If the connection is suddenly cut off, B can infer that A has taken some deliberate action to end the call; B can then use this assumption to make further inferences about A's intent.

---

[1] While there are some shared motivations and techniques for the preservation of ambiguity and the preservation of privacy, this proposition is *not* the same as simply saying that designers must consider privacy issues; see the discussion in the later section on "Ambiguity in Design."

[2] Garfinkel generally uses "account" as a noun, to mean a description that serves to lead others to see that one's actions make sense in context. In interaction, one strives to make one's actions "account-able" at all times – all parties should have an idea of what each of the others is saying and why. Our use of "account" as a verb is an attempt to capture this implicit multi-sided notion – if one is constructing accounts of one's actions for others, it must be true that the others continually account for one's actions.

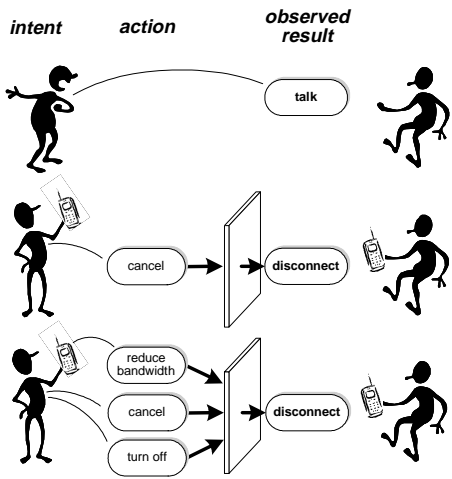

**Figure 1. Informal examples of ambiguity resulting from mediation and from multiple possible actors/actions.**

In more complex, "intelligent" forms of mediated communication in which an autonomous device may act on the user's behalf (Figure 1, bottom), not only is the exact action that resulted in the observed result unclear, but it may not even be clear which *actor* initiated the action. Person A's intent is doubly hidden from person B by a multiplicity of possible actions and possible actors. For example, say A and B have established an open audio channel (again, using an ideal service). If the open channel closes, person B cannot tell if (e.g.) the audio stream was terminated by software to reduce bandwidth consumption, the channel was deliberately selected for cancellation, or the handset itself was switched off. Person B also cannot conclude whether person A or their device took an action.

To summarize, it is harder for observers to tie observed results or actions to personal intentions and dispositions if there are many possible explanations. As mentioned above, this general idea comes up in many theories in various fields. For example, consider attribution theory in social psychology: "An action is seen as corresponding to a particular intention to the extent that other possible reasons for the action can be 'discounted'…the fewer unique effects the chosen behavior has, the less ambiguous it is just which effect the actor intended" [7]. However, similar ideas can also found elsewhere (e.g., in sociology, one can consider [12,15,31]; in linguistics, [5]). We need not "buy into" attribution theory, ethnomethodology, or other specific theories to be able to reason about these processes.

**Active Effects of Mediation**

Of course, the effect of mediated communication on accounting is not limited to that of the structure of the communication channel. Participants take an active role, going beyond observing and accounting for others' behavior – they try to influence how others are accounting for their own actions. Furthermore, the communication system itself can provide features that influence how participants account for each other's actions. The former topic is the subject of much social science research, so here we will restrict our attention to the latter topic.

There are two main strategies that have been used in communication systems in the past to alter information to reduce the degree to which people can make inferences from it. First, the communication system can change the *precision* of information. Omitting the information entirely is an extreme form of this strategy. However, one can be more selective, altering only particular aspects of the information; for example, if one does not want people to be identifiable in a video stream, one can selectively blur portions of the video images. Second, the communication system can change the *accuracy* of information. Falsifying the information entirely is an extreme form of this strategy. Reduced accuracy is not unthinkable in social interaction, as prevention of undesired inferences often involves deception of some kind, even between friends [8]. Products exist that are solely intended to actively mislead others about illicit behavior (e.g., recordings of background sounds to fake one's "surroundings" in phone calls [10]).

However, ambiguity, being about meaning, is a social construction and a designer of a mediated communication system can go farther than just processing particular data types or data streams according to the right set of rules. The point is that appropriate degrees of ambiguity can be provided through careful design. The idea is not merely to try to hide the truth, or to try to convince someone of an untruth, but rather to multiply the possible situations so that negative formulations can be avoided if mutually desired.

**DESIGN CASES**

The previous sections have left open the question of how we can design lightweight communication systems to provide appropriate resources for interactional ambiguity. Rather than trying to consider all relevant resources and possible designs, we will instead focus on the uses of ambiguity to facilitate the ability of users to be interactionally unresponsive. In this section, we first explain the problem and its importance. We then work through two illustrative design cases. The first case illustrates how empirical methods can be applied to assess the use of ambiguity in a communication system; as such, it is based primarily on an analysis of fieldwork of users using an existing commercial communication system, and it gives a relatively simple and grounded example of how a particular design succeeds in providing an appropriate kind of ambiguity. The second case is more complex and theoretical, illustrating how one might move forward from empirically-grounded insights and apply them in the design process. Motivated by fieldwork findings from the literature, we describe a design technique that facilitates a different kind of ambiguity and work out two design scenarios for concreteness.

Unresponsiveness has been largely unaddressed in previous work: research in lightweight communication systems has tended to focus on designs that allow interactions to *begin* with greater spontaneity, rather than on facilitating the ability to avoid, pause, or escape interactions as needed. However, many situations exist in which it is desirable to delay or avoid communication. People are constantly balancing demands on their attention. Even in face-to-face interaction, balancing the attentional demands of co-present parties and tasks can be difficult. The challenge is much greater as mobile technologies make communication with remote parties more pervasive.

The fundamental problem with unresponsiveness, however, is that interaction is an affirmation of a social relationship as well as communication [15]. For this reason, the closing sequence [30] of a telephone call between people with an established relationship often involves an affirmation that they will talk again (and perhaps even when this will happen). Consequently, a simple lapse contains an element of social insensitivity. Even in co-presence, an abrupt lapse in talk is a social signal (usually of rudeness) when it occurs between people who are not well-attuned to each others' behavior.

Our goal in this section, then, is to discuss how interactional unresponsiveness can be facilitated through appropriate ambiguity. We describe this at two levels. The first is the *conversational* level, i.e., managing responsiveness within a single interaction. (We use "conversational" in a loose, colloquial sense, meaning only a sequence of communications – textual messages, turns at talk, etc. – that are responsive.) At this level, we assume that users have the ability to reach each other in a given medium, and we consider how particular types of ambiguity may support delayed responses or spontaneous lapses in conversation. The second level is *associational* or relational, i.e., managing communicative access with others. At this level, users establish boundaries regarding who can contact ("associate" with) them in a given medium. For example, users make decisions about to whom they will give their telephone number.

In terms of the relationship between the conversational level and the associational level, one may broadly consider that conversational is within an interaction (or sequence of interactions) in a medium, while associational determines whether that interaction can occur. Our goal is to consider designs that support stories at both levels.

**Case I: Conversational Ambiguity and Push-to-Talk**
As just mentioned, unexplained incidents of unresponsiveness in face-to-face interaction and real-time mediated communication tend to have negative social implications. For example, when communicating through a full-duplex audio connection, such as a telephone call or an open audio channel, people generally feel they must respond promptly when someone speaks to them (see, e.g., [1]). However, this is not true to the same degree in all forms of mediated communication. In this subsection (which expands substantially on a brief discussion in [37]), we discuss our experience with push-to-talk audio provided by cellular radios. Observed conversation in push-to-talk audio contained delayed responses. We first discuss the phenomenon of response delays and then discuss how they were related to ambiguity. We then relate this discussion to previous studies of other mediated communication technologies to try to separate design issues from the particulars of the specific technologies.

*Release from Responsiveness*
These findings are based on a study conducted in June 2002 [37]. We provided a close-knit social group of seven college students with cellular radios and observed and interviewed them regarding their use of the technology.

Cellular radio service is push-to-talk, similar to walkie-talkies or other radio services (however, it is provided through a cellular network rather than through direct radio signaling). A number of factors distinguish push-to-talk interaction from telephone interaction. Most relevant to our discussion is the fact that audio is only transmitted by a participant when they take the explicit action of pushing and holding a button. More specifically, if person A wishes to speak to person B, they press a button and begin speaking; person A's utterance emanates from person B's device as it is produced, with a network delay identical to that of a mobile phone call. Further, the half-duplex cellular radio channel inherently prevents overlap. By contrast, in a full-duplex channel such as a telephone call, both parties transmit continuously during a connection, e.g., a participant who is talking can hear audio from the other participant such as laughter or a baby crying in the background.

In interviews, participants reported that they did not feel they needed to reply immediately when someone spoke to them via the cellular radio.

> Kelly: "[I feel like I have to answer if somebody says something to me] but not immediately. I can do it on my own time… if I'm like busy or something like that, and then when I get a chance I'll be like, what did you say, what do you want?"

To see an example of this in practice, consider the following transcript. During this excerpt, Julie was at home while Todd was at work. Julie and Todd had had several recent conversational exchanges prior to line 1; the last exchange was 56 seconds ago. Prompted by a conversation with a co-present party, Julie asked Todd about communication technologies he uses while at work. ("IM" refers to instant messaging, a textual medium that enables users to set up near-synchronous communication

sessions in a lightweight manner [27,34]. "AIM" refers to a specific IM system, AOL® Instant Messenger™.)

**Excerpt 1:**
1  Julie:  Why don't you use IM at work?
2         ((pause of 3 seconds))
3  Todd:  What?
4         ((pause of 0.2 seconds))
5  Julie:  Why don't you use IM at work?
6  ⇒      ((pause of 3 minutes and 13 seconds))
7  Todd:  Because they don't have AIM here.
8         ((pause of 1.4 seconds))

Note that in this case, a large pause occurs without apology, explanation, or recrimination within a question-answer sequence (marked by ⇒ in line 6). Consider how odd this would be in typical face-to-face or telephone conversation. Because she does not immediately pursue an answer, Julie is plainly accounting for Todd's pause – so long as to practically constitute a lapse in talk – as being caused by some acceptable factor.

We observed several important types of ambiguity relating to the cellular radios. Here we focus on two that are relevant to reduced accountability to respond quickly.

First, due to the push-to-talk nature of cellular radio transmissions, there was ambiguity about the recipient's activities. As with most types of audio communication, background noise can be heard while a participant is transmitting. However, because the channel is one-way and because users only transmit when they press and hold a button, information did not "leak through." For example, a transmitter could not hear whether a remote recipient was engaged in activities such as conversing with other co-present people or typing. (Contrast this with most audible media such as telephones and audio spaces which are full-duplex so such activities can easily be heard by the other participant(s).) Therefore, as far as transmitters knew, a recipient might be doing anything from having an argument with their boss to playing a computer game to politely attending to the transmission.

> Erica: "I understand to wait if I'm talking to anybody till they're free and stuff [if they don't answer]… I tried to message Julie earlier but it wasn't working and I figured she was probably at work, busy in a meeting or something."

Again, as in Excerpt 1, we see how participants tended to account for lapses in positive ways. Of course, non-responsive participants may in fact have been rudely ignoring messages, but the ambiguity allowed participants to construe reasons that were socially acceptable.

Second, several factors resulted in ambiguity about whether or not a transmission was received. Although the cellular radios were mobile, participants were not always within hearing distance (e.g., they might have left the handset on a desk while in the bathroom). Since the medium was not persistent (reviewable [6], recordable [20]), transmissions sent during such periods were lost. Additionally, the medium was not entirely reliable, as transmissions were occasionally dropped by the network.

*Discussion*
The flexibility to reduce responsiveness without causing social difficulties can be quite valuable. This raises an obvious design question: are there simple principles for affording the kind of conversational ambiguity described above? In fact, the fieldwork on lightweight communication systems has identified a number of system design factors that appear to relate to reduced responsiveness. However, while these are likely to be useful in guiding design in a heuristic manner, predicting the final outcome may be more difficult.

The fieldwork on IM suggests two such factors. The first relates to a lack of "leak-through" (similar to that noted in cellular radio). In particular, there is the notable finding that recipients report a feeling of *plausible deniability* [27]. That is, recipients rely on the sender's lack of information to excuse a lack of responsiveness: "[T]he sender generally doesn't know for certain whether the intended recipient is there or not. As a result, failing to respond is not necessarily interpreted as rude or unresponsive" [27]. Note that this is not restricted to the initiation of an interaction. In ongoing interactions, replies to messages may be delayed, or interaction may (more rarely) lapse without explanation [27,34]. The person waiting at the other end simply has to assume that the other participant had some reason to stop responding because they have little visibility into the other participant's environment. (Even if positive presence information is available, senders do not always assume that the recipient is necessarily able to reply and therefore obligated to respond.) It is remarkable to find similar ambiguity-related behavior in both IM (a desktop, text-based system) and cellular radio (a mobile, audio-based system). The second relates to reviewability. It is suggested that message persistence makes it safer to delay responses in IM – the message is a reminder to respond, and both the sender and recipient are aware of this [27].

The fieldwork on text messaging suggests a third factor, one that relates to mobility. SMS text messaging enables users to send short, asynchronous messages between mobile phones. It is reported that recipients feel pressure to reply "immediately" [22] (though timescales vary) and, more importantly, to explain any delays in replying [22,23,33]; it has been suggested that recipients may feel this pressure because the sender is aware that the handset is carried "at all times" [22,23]. (Note that reviewability plays an implicit role here as well.)

While these factors are plausible, it may prove difficult to reason about them in aggregate. From Table 1, it is hard to identify individual factors or combinations of factors that explain the reported feelings of plausible deniability and/or reduced responsiveness. For example, since the factors in Table 1 have been (re)formulated so that each increases the expectation of responsiveness, one might hypothesize that media with fewer of these factors are more likely to afford reduced responsiveness. The difficulty is that SMS and IM would be predicted to be more similar than IM and cellular radio, which has not been the case.

| medium | m. phone | SMS | IM | c. radio |
|---|---|---|---|---|
|  | audio | text | text | audio |
| "leak-through" [27,37] | X |  |  |  |
| *not* reviewable [27] | X |  |  | X |
| mobile [22,23] | X | X |  | X |
| reported "deniability" |  |  | [27] | [37] |

**Table 1. Some factors relating to reduced responsiveness.**

This is hardly an exhaustive analysis, and there are many more factors (such as media affordances, costs and constraints [6,13,36]) that could be investigated. However, there are at least two reasons to suspect that reasoning about a medium in isolation will prove difficult.

First, the factors and the phenomena of interest are somewhat soft. In particular, the reported feelings of reduced responsiveness may be more a matter of how people orient to the medium than how they actually behave in the medium. For example, people do actually explain some of their response delays in IM, even though they extol the ability it gives them to be unresponsive [34].

Second, various factors relating to emergent practice are relevant. Studies of email [13] and IM [34] suggest that senders who do not receive a response to a communication appear to account for unresponsiveness in a way that is consistent not only with their understanding of the medium but also with their knowledge of common practice and conventions in that medium. Such practices and conventions may even have been influenced by other media; multiple participants in the cellular radio fieldwork made interview comments of the form, "It's kind of like IM over the phone" [37], offering the comments as a reflexive basis for their unresponsiveness in the cellular radios. A medium constrains, but does not determine, people's attitudes and practices around responsiveness [13].

At this point, fieldwork has revealed the presence and utility of conversational ambiguity in various media. However, while simple guidelines may be helpful in the design process, much more work needs to be done before we can predict how conversational ambiguity will operate in a given medium. The effectiveness of a design with respect to conversational ambiguity should be evaluated through trial use.

### Case II: Associational Ambiguity and Leases

We now turn to the notion of unresponsiveness at the associational level. In the previous subsection, we focused on an analysis of an existing design. In this subsection, we structure the discussion around the design process and how ideas related to ambiguity might be worked into it. We begin by reviewing existing support for access in mediated communication systems. We then introduce the notion of leases to allow parties to negotiate access in mediated communication systems, showing how they provide resources for important social processes. After going through two illustrative design exercises, we conclude with a discussion of some broader implications.

*Social Difficulties Caused by Existing Access Mechanisms*
Today, most widely-deployed personal communication systems can be characterized as using static mechanisms that rely on social mediation for access control. That is, access to a given communication channel is enabled by sharing a static piece of *address* information, such as a phone number; possession of a given user's address implies the (technological) ability to contact that user at any time. Potential contactees are expected to control who has their address, and contacters are expected to behave in a reasonable manner. Such social mediation is often difficult. The desire for contact is often asymmetric and it can be socially awkward to refuse to provide contact information [25]. Participants in our push-to-talk fieldwork complained that this static access does not meet their changing needs:

> Julie: "But like you give someone your cellphone number once and they think it's like how they can always get ahold of you."

The key issue here is that social relationships (including work relationships) evolve, and as they change, the desire of one person to be contacted by another often changes.

Technologists often have the idea of putting the contactee in total control by using access tokens that can be revoked or expired (e.g., [21]); early examples include telephone proxy services that allow one to receive live responses to personal advertisements without giving out "real" contact information (for example, see U.S. Patent 4,847,890). This is plausible if the right power relationships are in place between the parties. However, since it is just as much of a *unilateral* predeclaration of relationship limits as a refusal, it is no less problematic in more general settings – as with refusal, the entire onus of access denial is on the contactee.

Faced with these issues, users frequently resort to other strategies, such as avoidance. Some strategies greatly reduce the utility of the personal communication system. For example, users may change their address information, as when a user changes their phone number to avoid a persistent caller. More incremental strategies often have drawbacks as well.[3] For example, call screening requires the user to modify their call-answering behavior relative to all callers for an indefinite period in order to cope with a single unwanted caller. Ultimately, users may even abandon a technology entirely.

---
[3] An interesting incremental strategy known as "cyber-ditching" has become associated with the youth-oriented Microsoft 3° service [38] and shows how much effort some users put into avoidance. 3° provides a notion of a "group" that is akin to a shared desktop workspace for socializing. The designers report that users have investment in their screen names and so prefer not to change them, but since groups are invitation-only, a subset of the original members can effectively "revoke" the access of other members by (1) creating a new group and (2) agreeing not to re-invite the excluded members.

*A Proposal: Leases for Negotiating Access*

Users' strategies for managing access in these static systems are often cumbersome and/or have uncomfortable social consequences. In the rest of this subsection, we discuss novel leasing mechanisms that offer users more resources for managing who has access to them.

The approach we consider here is a service to provide renewable *leases* of temporary duration. The concept of a resource lease is often used in distributed systems applications such as file system cache consistency [17]. In such applications, a *resource manager* (RM) leases access rights to a *resource consumer* (RC) for an explicitly negotiated time period. At the end of this time period, the RM revokes the RC's rights (i.e., the lease *expires*) unless the RM and RC go through another negotiation and come to mutual agreement on a lease *renewal*. The key aspects of a leased resource, then, are (1) initial access is by negotiation between the RM and RC; (2) in the absence of appropriate subsequent interaction, access expires without further action on the RM's part; and (3) renewal occurs only by negotiation, i.e., if the RM and RC both act and concur. Hence, leases are a technical means of capturing expressions of ongoing reciprocal interest between entities.

As a design exercise, we applied the notion of leases to access in mediated communication channels. In this design space, the RM and RC are human social actors instead of file system clients and servers. (In a bidirectional communication channel, each user may serve as the RM in one direction and as the RC in the other.) Initial setup of a lease might consist of transmitting a digital representation of a cryptographically-secured capability. If renewal conditions are met (e.g., both parties contact each other frequently), access continues. Otherwise, if expiration conditions are met (e.g., the passage of a specified period of time), the lease expires. When a lease expires, the RM is notified, at which point the RM may renew the lease, reallocate the lease to someone else, or leave the lease unassigned. The RC is not notified about the specific condition causing the expiration or about the precise action the RM has taken.

The use of leases in a social context is, of course, mainly intended to afford several kinds of ambiguity, particularly those relating to actors, actions and intentions (Figure 1). For example, lease expiration will in some cases be due to a mutual failure to act, which distributes the social onus. Where the RC has clearly acted, expiration may have occurred because the RM explicitly chose not to renew (perhaps by clicking "I don't want to talk to this person anymore" in a dialog box) or the RM may have simply failed to take some action (perhaps being too busy or distracted to renew). The RC will not know if the RM has taken the more offensive explicit action or committed the less offensive action of simply "forgetting." (And, in fact, the RM can offer "I forgot" as an explanation if asked in either case.) Even if the actions are known, the intentions of that party may not be clear. As a result, leases inherently provide resources for explanations and stories that are not available with static access mechanisms.

A few caveats are in order concerning some simplifications we will make for reasons of clarity. The first concerns terminology. We will continue to refer to leases as such, to the party issuing the lease to be contacted as the RM, and to the person using the lease to contact the RM as the RC. It should be understood that these exact terms and abstractions would not necessarily be presented to end-users. The second concerns the role of leases when there are multiple communication media available. We discuss leases here as if they correspond to a given medium, but the idea can be applied in a variety of ways (many of which are relatively straightforward) that integrate multiple media.

*Exercise I: Scarcity as Resource for Stories*

We now turn to the first of our design exercises, which was inspired by the observation that people frequently use externally-imposed limitations – whether real or fictitious – as resources for face-work. Such limitations include money (students often have financial constraints which they discuss explicitly and use as socially acceptable excuses) and time (as when people say they do not have time to talk or to go to dinner). Personal communication service plans are a rich source of explanations, as they often impose limitations that are fairly arbitrary from the subscriber's perspective.[4] Teens and young adults with limited funds often explain that they cannot talk on the phone by saying they are out of minutes.

This observation can be translated straightforwardly into a design in which an RM's service subscription includes a fixed number of leases. This may be the same for all users of the service, or may vary depending on the price of their subscription; in either case, the key social resource is that *the limits appear to derive from a technological or economic basis*, as they often do with today's mobile phone calling plans. The limited supply of leases offers a plausible explanation of the intent behind a failure to renew: that the RM had to reallocate the lease to someone else with whom they had an urgent need to communicate.

Scenarios were a major part of the exercise; one is excerpted here for illustration. Social science research providing taxonomies of excuses and justifications [31] and politeness strategies [5] can be very helpful in grounding the outcomes of such scenarios.

**Excerpt 2:**
Two high school girls, Monica and Christina,…are assigned to work together on a class project. Christina (RC) asks Monica (RM) for her push-to-talk number so that they will be able to coordinate their work using the cellular radio… [Afterward,]

---

[4] Mobile phone service plans in the United States frequently impose limits on the number of minutes a subscriber may use during a month, with severe charges for exceeding the limit. The plans are often quite complex, imposing differential charges and minute limitations depending on whether the subscriber is contacting a subscriber to the same service provider, or a member of a pre-defined group of friends and family, etc.

Christina has an interest in continuing social interactions with Monica… Monica [does not and] begins to keep her push-to-talk cellular radio turned off most of the time. Alternatively, consider how this situation might play out with a lease. Monica allows the lease to expire… Monica can offer a plausible excuse such as "Oh, sorry, I needed that lease to give to my project partner for my history class. I only have five leases because my parents won't pay for more. I'll try to get one for you soon."

Whether or not the stories seem "transparent" is not (necessarily) the point. Such explanations are often transparent in that sense; preservation of face requires credibility and the willingness to accept an explanation, not actual belief [16].

It is important to note that the design process must continue to consider social affordances and user interaction issues other than associational ambiguity. For example, externally-imposed scarcity has lead to the appropriation of systems as tools for defining and negotiating the status of social relationships. The limited capacity of mobile phone address books makes them one such tool, as when users inspect and even edit each other's address books [19]; this is particularly true for the teen and young adult age groups, who are in a phase of life in which defining relationships through indirect mechanisms is appealing [18,19]. Leases could be similar resource, and additional scenarios have considered possible practices arising around the initial negotiation of leases; renewal of leases; and even, in some cases, the dramatic cancellation of leases.

*Exercise II: Context-awareness as Resource for Face-work*
Our second design exercise was inspired by ongoing research in context-aware computing. For example, one could base lease expiration on contextual data, such as the degree of contact between two people for a certain period of time – if the lease was unused, if it was used too much, if one party failed to return the other party's calls, if the parties were not in each other's physical presence, etc.

This type of lease increases ambiguity of actor, action, and intention because explicit lease-related actions are taken only by the system, and the system is a "black box." In particular, if the system aggregates information about multiple users' implicit actions (e.g., physical proximity), it can be difficult for a given user to draw inferences about what actions have been taken by another user. In many cases, a participant cannot rule out the idea that their own actions resulted in the expiration. Note that, in contrast to situations in which mysterious contextual inference negatively impacts usability [4], this is one in which "black box" contextual inference may have advantages.

As in the previous exercise, scenarios helped in thinking out possible outcomes of ambiguity-related strategies as well as technological design issues.

> **Excerpt 3:**
> Michael and Carrie are both in their thirties. They meet at a bar and Michael (RM) gives Carrie (RC) his phone number. They date briefly… Michael becomes disinterested in Carrie and… starts screening his calls. [This] affects his telephone behavior with all callers. Alternatively, consider how this situation might play out with a lease. When they first meet, Michael "beams" a lease from his phone to Carrie's using Bluetooth, one that will expire if two weeks pass without them being physically proximate (as measured by their phones recognizing each other's Bluetooth identifiers).… When they stop seeing each other…the lease expires… Carrie must consider many possible explanations. [S]he may decide that they are jointly responsible because they have not been in frequent contact. She might even choose to hold herself primarily responsible, thinking that if she had kept her date to meet Michael for coffee last week, the lease might have been renewed…

Again, both parties' choice of explanation is not just about facts, but also about the degree to which they wish to preserve face.

*Discussion*
With respect to the specific idea of leases, we have provided illustration rather than definitive guidance. The designer must make a number of strategic choices for any particular implementation, and the intended context for a given system (user population, situations, etc.) provides key factors in navigating the design space. We intend to follow an iterative design process with these leases, designing and conducting user studies to learn more about which mechanisms are most effective in which situations.

In this subsection, and in particular in the two design exercises, we have attempted to bring out several points that have application beyond leases. First, we have tried to illustrate that socially useful forms of associational ambiguity can be located by examining existing practice and how they might be targeted in the design process. Second, while the general idea of adding mechanisms to afford such ambiguity may seem to burden the user with additional effort, it should not be dismissed as the specific mechanisms can have significant benefits. Such mechanisms may be less burdensome than the workarounds that users otherwise employ (such as avoidance), and they may have additional social benefits (such as resources for defining social relationships). Third, we have tried to reinforce a point made in the previous design case, which is that while ambiguity mechanisms can provide the space for stories and face-work, the rest is up to the participants. Managing access still requires skill in story-telling, social negotiation, and managing power relationships [18].

**AMBIGUITY IN DESIGN**
In the previous section, we observed that designing to provide "appropriate" resources for ambiguity in communication systems is not always a straightforward proposition. Other areas of research in computer science and the social sciences address the question of ambiguity in communication, and it seems plausible that they could provide guidance. We have touched on many of these already, and in this section, we provide pointers into some additional relevant areas of research, making caveats as necessary about their applicability.

Discussion of ambiguity in the general design literature is generally limited to attempts to cope with it on the "human" side of human-computer interaction, such as the engineering of systems that must interpret human gestures [26]. When ambiguity is discussed on the "computer" side of HCI, it is generally considered to lead to a frustrating user experience and therefore held to be undesirable. There are forceful arguments that the actions of computer systems should be made intelligible to users [9]. It is, of course, a design truism that a system should "let the user know what is going on" (even though not all systems do so); this advice is held to be even more important in systems which cannot completely hide system complexities [9] or that make inferences about users' intentions [4].

There are, however, a number of perspectives that argue for a role for ambiguity in design:

*Critical design*. In a recent paper [14], Gaver *et al.* draw on examples from art and critical design to illustrate the uses of ambiguity in developing "engaging and thought-provoking" artifacts. They provide a taxonomy of ways in which ambiguity can be made relevant to interactive systems design that can serve as a resource for expanding our discussion of the active effects of mediation.

*Human communication*. While the focus in this paper has been on the process of designing to provide resources for ambiguity, a substantial body of research on human-human interaction can be mined for insights on how the resulting story-telling and face-work might unfold. At least three areas can be seen that would be very useful in scenario construction. First, researchers in areas such as social psychology, communication, and sociolinguistics have examined how people communicate ambiguously for various purposes. For example, there have been detailed examinations of the goals and strategies that people apply in face-work [5,16], equivocation [3], and related forms of deception [8] during social interaction. Second, there has been research on how the technical characteristics of different communication media interact with communicative ambiguity. Some was mentioned in the first design case (e.g., [13]), but there are a number of theories and studies on the ways in which different media affect how the content of messages is constructed and interpreted (see [35] for a recent survey). Third, there has been research on how people choose between various media that are available to them. For example, studies of people's media choices to facilitate deception [8,20] suggest that different media may be favored based on the users' beliefs about the likely outcome (including others' interpretations) of using them.

*Privacy*. While much work in privacy can be summarized as information hiding plus access control, recent HCI work has applied privacy regulation theory from social psychology [29]; this theory describes privacy as a dynamic and highly contextual boundary regulation process in which technological mediation plays a role. Research on systems for providing activity awareness and presence information also shares many of the concerns discussed in this paper, and much of it is quite sophisticated – for example, Erickson and Kellogg's notion of social translucence [11]. All of these outlooks are useful for design, but perhaps less applicable than our framework at the "micro" level at which one analyzes individual conversations.

We close this discussion by noting an important difference between frameworks that consider "information" (privacy, awareness,…) and a framework that considers "stories" and face-work. In the former, "information" is usually defined in a factual way, whereas in the latter, the most important element of successful face-work is not its truth *per se* but that it is accepted by the participants. Ambiguous situations often work as part of a story, not because they are factually convincing, but because they leave room for a face-saving formulation of the situation – and the ongoing relationship is more important than clarity.

**CONCLUSIONS**
In this paper, we have touched upon the complex relationships between ambiguity in social interaction, the process by which participants in social interaction account for each others' actions, and the relevant technological aspects of mediated communication systems. In particular, we have highlighted the notion that accounting for others' actions is a continual process, both in individual interactions and in relationships; an analysis of a system design with respect to ambiguity should consider not only what information passes through the system, but what the effects are on a user's ability to account and, when necessary, to execute face-work.

This discussion can be expanded in many ways. While the discussion has been framed quite generally in some sections, the design cases presented here are quite specific – they relate to particular interactional difficulties that occur in lightweight communication systems. Plainly, there is room for a great deal of theoretical and design-oriented generalization. There is also an enormous amount of empirical work that needs to be done – increasing our understanding of phenomena such as plausible deniability, evaluating the effectiveness of leasing in practice, etc.

However, what we have attempted to do is to fortify the dialogue between designers of mediated communication systems and social scientists. Rather than simply having social scientists analyze users' behavior in communication systems that have been designed with relatively arbitrary social affordances, we hope that future design discussions might include explicit consideration of how users make sense of their interactions and relationships.

**ACKNOWLEDGEMENTS**